\documentclass{scrartcl}

\usepackage[T1]{fontenc}

\KOMAoption{abstract}{true}
\KOMAoption{twocolumn}{off}
\KOMAoption{parskip}{half}
\recalctypearea

\usepackage{PackageManager}
\usepackage{hyperref}
\hypersetup{
	colorlinks  = false, 
	linktocpage = true, 
	breaklinks  = true, 
}

\title{\LARGE
A moving horizon estimator for \\ aquifer thermal energy storages
}
\date{April 2025}

\author{
    Johannes van Randenborgh and Moritz Schulze Darup
    \thanks{
        J. van Randenborgh and M. Schulze Darup are with the Control and Cyberphysical Systems Group, Faculty of Mechanical Engineering, TU Dortmund University, Germany
        {\ttfamily\small \{<first name>.<last name>\}@tu-dortmund.de}.
        This manuscript is accepted for publication at the European Control Conference 2025 in Thessaloniki, Greece.
    }
}

\titlehead{
        $\copyright$ 2025 EUCA.
        This work is a preprint of a contribution to the European Control Conference 2025 with the DOI: \href{https://doi.org/10.23919/ECC65951.2025.11186974}{10.23919/ECC65951.2025.11186974}.
}

\begin{document}
\maketitle
\begin{abstract}
Aquifer thermal energy storages (ATES) represent groundwater saturated aquifers that store thermal energy in the form of heated or cooled groundwater.
Combining two ATES, one can harness excess thermal energy from summer (heat) and winter (cold) to support the building's heating, ventilation, and air conditioning (HVAC) technology.
In general, a dynamic operation of ATES throughout the year is beneficial to avoid using fossil fuel-based HVAC technology and maximize the ``green use'' of ATES.
Model predictive control (MPC) with an appropriate system model may become a crucial control approach for ATES systems.
Consequently, the MPC model should reflect spatial temperature profiles around ATES' boreholes to predict extracted groundwater temperatures accurately.
However, meaningful predictions require the estimation of the current state of the system, as measurements are usually only at the borehole of the ATES.
In control, this is often realized by model-based observers.
Still, observing the state of an ATES system is non-trivial, since the model is typically hybrid.
We show how to exploit the specific structure of the hybrid ATES model and design an easy-to-solve moving horizon estimator based on a quadratic program.
\end{abstract}

\section{INTRODUCTION}\label{sec:introduction}
The attendees of the 28th Conference of the Parties emphasize the end of the fossil fuel era to mitigate raging climate change~\cite{UNFCC.Dec.2023}.
Considering the greenhouse gas emissions related to building operation and, more precisely, its heating, ventilation, and air conditioning (HVAC) technology, environmentally friendly HVAC solutions must be deployed globally to contribute to the attendees' aims~\cite{InternationalEnergyAgency.2023}.

A promising and ubiquitous approach is to harness - currently just dissipating - excess thermal energy with storage solutions.
In general, one must highlight that storages for building climatization are concomitantly in use with other, mostly fossil fuel-based, HVAC technology.
Thus, the storage's operational aim is a dynamic and maximal contribution to the thermal energy demand~\cite{Lee.2013}.

For this publication, the considered storage solution comprises aquifer thermal energy storages (ATES) and a cocurrent heat exchanger (HX).
Figure~\ref{fig:ATES-principle} illustrates the ATES and HX.
Further, it indicates concomitant HVAC systems with two auxiliary power units (APU), which can represent any HVAC technology.
The APU and the building are grayed out and given by a dotted line as they are not considered here.
The storages pre-heat or -cool the building's process fluid before it flows into the APU.
In this setting, the APU's energy consumption may be economically lowered by the ATES system \cite{Stemmle.2024}.

\subsection{ATES system operation and its control related challenges}\label{sec:ates-system-operation-and-its-control-related-challenges}
In principle, ATES systems store heat and cold in groundwater saturated aquifers (at least one for heat and cold), whereby thermal energy, in the form of heated or cooled groundwater, is simultaneously injected and extracted.
In winter, thermal energy from the warm storage passes via an HX to the building.
There, the groundwater cools and flows into the cold aquifer (see Figure~\ref{fig:ATES-principle}).
Storing the cold in the ground leads to local temperature changes in the soil.
During summer, the cold aquifer releases and delivers the cooled groundwater, which heats and runs into the warm aquifer.
Given that no thermal energy is demanded, there is no pump flow between the storages and, thus, no thermal energy exchange between the ground and building.
The operational modes of ATES systems are heating, inactivity, and cooling.

The use of ATES systems depends, among other things, on the correct and complete integration of the system into the control strategy of the whole building~\cite{Fleuchaus.2020b, Bloemendal.2022}.
In this context, \cite{Drgona.2020} discuss general model predictive control (MPC) strategies.
MPC, as a model- and optimization-based controller, has the advantage of predicting future states of the building and using these predictions to determine an optimal system input, e.g., flow rates, valve positions, etc.
Of course, the performance of MPC is, i.a., limited to the prediction accuracy of the embedded model and the prediction horizon \cite{Hoving.2017}.
This aligns with the given challenges by \cite{Drgona.2020} for a worldwide deployment of MPC for buildings, which is the design of a "[u]ser-friendly, control-oriented, accurate, and computationally efficient [...]"~\cite{Drgona.2020} building model.

Referring to ATES systems, such a model should predict the extracted groundwater temperature of the storages \cite{Hoving.2017}.
The MPC uses these temperatures to correctly estimate ATES' energy contribution to the building's energy demand.
Consequently, the model must capture, at least to a certain extent, the spatial distribution of the ground's temperature.
The shortcoming of temperature sensors in the ground, due to high drilling costs, promotes state estimators to a crucial element in the control loop.
Every deviation of the MPC's predicted power supply must be compensated by the APU.
This may be considered expensive in monetary terms and greenhouse gas emissions.

\subsection{Known MPC and state estimation approaches for ATES}\label{sec:known-mpc-and-state-estimation-approaches-for-ates}
Literature describes conservative~\cite{Vanhoudt.2011} and advanced MPC control solutions for ATES systems.
Given the context of this publication, the following paragraphs cover a brief review of MPC for ATES systems only.
To the knowledge of the authors, literature's MPC models typically assume spatially homogeneous temperatures for the storages, which cannot properly grasp ongoing physical effects in the ground.
As a consequence, not all unveil the necessity of state estimators for spatial temperature distributions in the ground.
For the sake of completeness, the pioneering work is still referenced.

A simple thermal energy-based first principle linear model for ATES systems is presented by \cite{Rostampour.2016c}, which considers heat loss by a time-invariant constant and assumes a spatially homogeneous storage temperature.
Building on that, \cite{Rostampour.2017} present an MPC approach for smart thermal grids including several clients (buildings) and ATES systems.
The MPC constantly monitors the ATES' power delivery.
Perfect mixing of injected enthalpy is assumed for the storages, which results in spatially homogeneous temperatures.

Hoving et al. \cite{Hoving.2017} use data from an existing ATES system to fit a model including two ATES, heat pump, and district heating.
The authors assume constant ATES temperatures for the warm and cold storages.
The MPC focuses on achieving a balanced operation, which is required by legislation.
To avoid the degradation of the cold storage, a sub-component of the HAVC system, an air handling unit, can additionally add cold to the storage.

To the knowledge of the authors, the first publication that describes a state estimator for ATES systems is presented by \cite{vanRandenborgh.2024}.
Increasing the MPC model's physical fidelity, the authors propose to approximate the ATES system by a piecewise (or hybrid) nonlinear surrogate model.
This approach keeps track of the spatial temperature distribution in the storage.
Based on the surrogate model, the authors derive by linearization a piecewise affine (PWA) MPC model, which results in a mixed-integer quadratic program (MIQP).
The states of the MPC model are the spatial temperature profiles, which are initialized with a state estimate determined by an unscented Kalman filter (UKF).

\subsection{Contributions}\label{sec:contributions}
The design and performance of the state estimator are crucial for the success of MPC for ATES systems.
Wrong estimations influence the MPC's control input and, by this, the desired strategy of the MPC scheme.
Ideally, a state estimator combines both properties: stability and optimality \cite{Rawlings.2022}.
Whereas \cite{vanRandenborgh.2024} have shown that the proposed UKF is stable, the optimality of the state estimations is questionable.
By design, the UKF does not obey physical (system) constraints, which lowers its estimation accuracy, e.g., see \cite[Example 4.39]{Rawlings.2022}.
Consequently, the UKF may estimate temperatures lower than the freezing limit.
Furthermore, the UKF suffers from the ``short horizon syndrome'', which aims at the consequences of considering only the latest measurement and not a sequence of past measurements \cite[Ch. 4.4]{Rawlings.2022}.
Indeed, large initial state errors (which may occur after a sensor malfunction) for the UKF result in inaccurate estimations and may destabilize the filter during operation.

In this publication, a moving horizon estimator (MHE) for ATES systems is presented, which includes physical constraints and a sequence of past measurements.
To tackle the (general) computational burden of MHE for piecewise systems, e.g., solving a mixed-integer quadratic program \cite[Ch. 4.D]{Bemporad.2000}, we exploit the structure of the nonlinear piecewise ATES model and propose a computationally cheap estimator design in the form of a quadratic program.
For this, the surrogate ATES system model by \cite{vanRandenborgh.2024} is extended.

\begin{figure}[t]
    \centering
    \includegraphics[trim={0cm, 11.55cm, 0cm, 0.2cm}, clip, width=\linewidth]{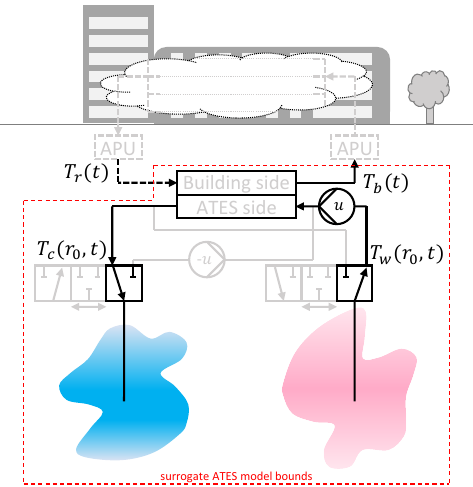}
    \caption{
        Schematic illustration of a building with an ATES system in heating mode comprising one warm (red) and cold (blue) aquifer, an HX, two APUs, and two groundwater pumps.
        Components illustrated by dotted lines are not considered by the ATES surrogate model.
        $T_{\mathrm{w}}$ and $T_{\mathrm{c}}$ refer to the temperature of the warm and cold storage.
        }
    \label{fig:ATES-principle}
\end{figure}

\section{PRELIMINARIES}
This section briefly introduces MPC, piecewise systems, and MHE, and it is primarily meant to introduce the reader to the notation.
A profound introduction to MPC can be studied with, e.g., \cite{Borelli.2017} or \cite{Kouvaritakis.2016}, and a thorough introduction to state estimation is given by, e.g., \cite{Simon.2006} or \cite{Astrom.2009}.

The discrete-time domain and an indicator of ``current'' time (``experienced'' by the system) must be introduced first to explain the chronological order of the MPC and MHE algorithms.
The index~$k \in \N_{0}$ of $t_{k}$ highlights the discrete-time domain and links to the $k$-th multiple of the discretization step size~$\Delta t$; $t_{k} = k \Delta t$.
Here, $\N_{0}$ denotes the natural numbers including zero.
Further, the time that is ``experienced" by the system, the current time, is marked with the index~$k^{\circ}$ (for reference purposes only).

\subsection{Basics on model predictive control}\label{sec:MPC}
MPC's control input rests on the solution of an optimal control problem (OCP) that considers model-based state predictions and constraints on inputs and states.
As time proceeds, the MPC determines the control input by recursively solving the OCP every time~$t_{k}$.
The solution of the OCP delivers the optimal control input sequence~$\ub_{N}$, which is partly applied to the system.
The input sequence~$\ub_{N}$ holds the inputs~$\ub(t_{k}) \in \R^{m} \, \forall k \in \{k^{\circ}, ..., k^{\circ} + N - 1 \}$, which are at and ahead of current time~$t_{k^{\circ}}$.
After solving the OCP, the first element of the input sequence~$\ub_{N}$, $\ub(t_{k^{\circ}})$, is applied to the system.
The general OCP
\begin{align}
    & \min_{\ub_{N}} \Jf(\xb(t_{k^{\circ}}), ..., \xb(t_{k^{\circ}+N}), \ub_{N}) \label{eq:mpc} \\
    & \text{subject to} \nonumber
\end{align}
\vspace{-0.9cm}
\begin{align*}
    \xb(t_{k+1}) & = \ff_{\mathrm{mpc}}(\xb(t_{k}), \ub(t_{k})) \quad \forall k \in \{k^{\circ}, ..., k^{\circ}+N-1\} \\
    \xb(t_{k}) & \in \mathbb{X} \quad \forall k \in \{k^{\circ}, ..., k^{\circ}+N\} \\
    \ub(t_{k}) & \in \mathbb{U} \quad \forall k \in \{k^{\circ}, ..., k^{\circ}+N-1\}
\end{align*}
minimizes the objective function~$\Jf(\cdot)$ considering the dynamics~$\ff_{\mathrm{mpc}}(\cdot) : \R^{n} \times \R^{m} \mapsto \R^{n}$ of the state~$\xb(t_{k}) \in \R^{n}$ and constraints on state~$\xb$ and input~$\ub$.
$\mathbb{X} \subset \R^{n}$ and $\mathbb{U} \subset \R^{m}$ are convex sets.
Here, the index ``mpc'' of $\ff_{\mathrm{mpc}}(\cdot)$ refers to the MPC model.

\subsection{Piecewise and piecewise affine systems}
Piecewise systems generally comprise a nonlinear state~$\ff_{\mathrm{pw}}(\xb(t_{k}), \ub(t_{k})) : \R^{n} \times \R^{m} \mapsto \R^{n}$ and output function~$\hf_{\mathrm{pw}}(\xb(t_{k}), \ub(t_{k})) : \R^{n} \times \R^{m} \mapsto \R^{p}$, whose domains are divided in $s \in \N$ partitions (or sets) $\Xbc_{i} \subset \R^{n+m} \, \forall i \in \{1, ..., s\}$, such that
\begin{equation*}
    \begin{split}
\label{eq:PWNL-model}
    \ff_{\mathrm{pw}}(\cdot) & :=
        \begin{cases}
            \ff_{1}(\cdot) & \text{if } [\xb^{\top}(t_{k}), \ub^{\top}(t_{k})]^{\top} \in \Xbc_{1} \\
            \vdots \\
            \ff_{s}(\cdot) & \text{if } [\xb^{\top}(t_{k}), \ub^{\top}(t_{k})]^{\top} \in \Xbc_{s}
       \end{cases} \\
    \hf_{\mathrm{pw}}(\cdot) & := \begin{cases}
        \hf_{1}(\cdot) & \text{if } [\xb^{\top}(t_{k}), \ub^{\top}(t_{k})]^{\top} \in \Xbc_{1} \\
        \vdots \\
        \hf_{s}(\cdot) & \text{if } [\xb^{\top}(t_{k}), \ub^{\top}(t_{k})]^{\top} \in \Xbc_{s} . \\
    \end{cases}
    \end{split}
\end{equation*}
For all partitions~$\Xbc_{i}$ holds $\Xbc_{i} \cap \Xbc_{j} = \emptyset \, \forall i \neq j$, i.e., the sets~$\Xbc_{i}$ are not overlapping, and the union of all sets~$\Xbc_{i}$ complies with the state~$\mathbb{X}$ and input constraints~$\mathbb{U}$ \cite{Rawlings.2022, Bemporad.1999}.

Taylor expansions derive a PWA approximation of the nonlinear dynamics, where each function~$\ff_{i}(\cdot)$ and $\hf_{i}(\cdot)$ is transformed to affine dynamics,
\begin{equation*}
    \label{eq:PWA-model}
    \begin{split}
        \xb(t_{k+1}) & = \Ab_{i} \, \xb(t_{k}) + \Bb_{i} \, \ub(t_{k}) + \fb_{i} \\
        \yb(t_{k}) & = \Cb_{i} \, \xb(t_{k}) + \Db_{i} \, \ub(t_{k}) + \eb_{i} \, ,
    \end{split}
\end{equation*}
with the state matrices~$\Ab_{i} \in \R^{n \times n}$, the input matrices~$\Bb_{i} \in \R^{n \times m}$, and the state offset vectors~$\fb_{i} \in \R^{n}$ \cite{vanRandenborgh.2024, Borelli.2017, Bemporad.1999}.
The system output~$\yb(\cdot) \in \R^{p}$ is defined by the output matrices~$\Cb_{i} \in \R^{p \times n}$, the feedthrough matrices~$\Db_{i} \in \R^{p \times m}$, and the output offset vectors~$\eb_{i} \in \R^{p}$.

\subsection{Basics on moving horizon estimation for PWA systems}\label{sec:basics-on-mhe-for-pwa-systems}
The MHE aims to estimate the state~$\xb(t_{k^{\circ}})$.
To do so, the optimization problem (OP),
\begin{align}
    \min_{\xb(t_{k^{\circ}-M}), \nub_{M}} \sum_{k = k^{\circ}-M}^{k^{\circ}} & \|\nub(t_{k}) \|^{2}_{\Qb} + \| \omegab(t_{k}) \|_{\Rb}^{2} \label{eq:MHE-OP} \\
    & + \Gamma_{t_{k^{\circ}-M}}(\xb(t_{k^{\circ}-M})) \nonumber
\end{align}
\vspace{-0.8cm}
\begin{subequations}
    \label{eq:mhe-op-constraints}
    \begin{align}
        & \text{subject to} \nonumber \\
        \xb(t_{k+1}) & = \ff_{\mathrm{mhe}}(\xb(t_{k}), \ub(t_{k})) + \nub(t_{k}) \, , \label{eq:MHE-sys-state-dyn} \\
        \yb(t_{k}) & = \hf_{\mathrm{mhe}}(\xb(t_{k}), \ub(t_{k})) + \omegab(t_{k}) \, ,\label{eq:MHE-sys-output-dyn} \\
        \xb(t_{k}) & \in \mathbb{X} \text{ and }
        \nub(t_{k}) \in \mathbb{V}
        \label{eq:MHE-process-noise-constraints}
        \, ,
    \end{align}
\end{subequations}
is solved at time~$t_{k^{\circ}}$.
Note that the OP's constraints hold for all $k \in \{k^{\circ}-M, ..., k^{\circ} -1 \}$, and the solution yields the state estimate~$\xb(t_{k^{\circ}-M} | t_{k^{\circ}})$ of state~$\xb(t_{k^{\circ}-M})$ with knowledge at time~$t_{k^{\circ}}$ and the optimization variable~$\nub_{M}$, which is defined by $\nub_{M} = \left[ \nub^{\top}(t_{k^{\circ}-M}), ..., \nub^{\top}(t_{k^{\circ}}) \right]^{\top}$.
The OP assumes that the process noise~$\nub(t_{k}) \in \R^{n}$ is bounded by a polyhedral set $\mathbb{V}$ (see \eqref{eq:MHE-process-noise-constraints}).
$\omegab(t_{k}) \in \R^{p}$ denotes measurement noise, and the index ``mhe'' indicates the specific MHE model.
A corresponding measurement sequence with a finite time horizon~$M \in \N_{0}$ parameterizes the OP.
The state estimate~$\xb(t_{k^{\circ}} | t_{k^{\circ}})$ of state~$\xb(t_{k^{\circ}})$ is the final element of the state trajectory that starts with the state estimate~$\xb(t_{k^{\circ}-M} | t_{k^{\circ}})$ from the OP.

The objective function of the OP has three tasks.
First, the OP minimizes the process~$\nub(t_{k})$ and, second, the measurement noise~$\omegab(t_{k})$, which is implied by the equality constraint~\eqref{eq:MHE-sys-output-dyn}.
Third, the OP minimizes $\Gamma_{t_{k^{\circ} -M}}(\cdot) : \R^{n} \mapsto \R$, which can be considered as ``initial penalties'' that substitute missing system information from time~$t_{0}$ to $t_{k^{\circ}-M}$ \cite{Rawlings.2022}.

When $\ff_{\mathrm{mhe}}(\cdot)$ and $\hf_{\mathrm{mhe}}(\cdot)$ are piecewise systems, the OP transforms to a mixed-integer program and is computationally hard to solve (see, e.g., the deadbeat observer for PWA systems \cite[Ch. 4.D]{Bemporad.2000}).
Accordingly, integer decision variables are needed to determine the active partition of a piecewise system in the mixed-integer program~\eqref{eq:MHE-OP}.

Further, \cite{FerrariTrecate.2002} present assumptions that guarantee the stability of the MHE for piecewise systems.
The assumptions focus on the initial penalty~$\Gamma_{t_{k^{\circ}-M}}(\cdot)$ and the piecewise system.
The most important assumptions are: the gathered measurements must correspond to the PWA model at each partition~$\Xbc_{i}$, and the system must be incrementally observable.
Incremental observability can be verified with, e.g., \cite[Alg. 1 and 2]{Bemporad.2000} or \cite{Rawlings.2022}.
Furthermore, \cite{FerrariTrecate.2002} point out that the accuracy and convergence rate of MHE at the beginning of its operation depends on the initial penalties~$\Gamma_{t_{k^{\circ}-M}}(\cdot)$, and they show how to compute good initial penalties.

\section{SURROGATE ATES MODEL}\label{sec:surrogate-ates-model}
The surrogate ATES model serves as a fundamental basis for the MHE model, and its model boundaries are highlighted by the red dashed line in Figure~\ref{fig:ATES-principle}.
As already indicated in Section~\ref{sec:basics-on-mhe-for-pwa-systems}, the surrogate model is extended compared to \cite{vanRandenborgh.2024} and, now, encloses the outlet temperature at the building side~$T_{\mathrm{b}}$ of the heat exchanger.
To focus on the temperatures in the ATES and the energy exchange between the ground and the surface, the surrogate model allows to rudimentarily consider the building by its return temperature~$T_{\mathrm{r}}$ and a constant flow~$q_{\mathrm{b}}$ at the building side of the heat exchanger.
Further, thermodynamic effects of the building's piping system on the process fluid are neglected.

By assumption, the partial differential equation (PDE),
\begin{equation}
    \label{eqn:T_PDE_nonlinear}
    c_{\mathrm{a}} \frac{{\partial T(r,t)}}{{\partial t}} =  \frac{\lambda}{r} \frac{\partial}{\partial r} \left( r \frac{\partial T(\cdot)}{\partial r}\right) - c_{\mathrm{w}} \phi v(r,t) \frac{\partial T(\cdot)}{\partial r} \, ,
\end{equation}
governs the energy transport in the ATES, where $c_{\mathrm{a}}$ denotes the specific volumetric heat capacity of the aquifer, $v(r,t)$ the radial groundwater flow, $\lambda$ the conduction coefficient of the subsurface, and $\phi$ the porosity of the ground \cite{Anderson.2005, HechtMendez.2010}.
In fact, porosity~$\phi$ in the ground uncouples the speed of heat transport from the groundwater's velocity \cite{Bloemendal.2018}.
The solution of \eqref{eqn:T_PDE_nonlinear} yields the groundwater temperature~$T(r,t)$ over the radius~$r$ and time~$t$.
Note that \eqref{eqn:T_PDE_nonlinear} is one-dimensional and exploits the symmetry of the energy transport problem in the ground.
Finite volume methods (FVM) reformulate the PDE~\eqref{eqn:T_PDE_nonlinear} to a discrete-time and -space surrogate model (see, e.g., \cite{Schafer.2022} or \cite{Ferziger.2002}).

For the solution of \eqref{eqn:T_PDE_nonlinear}, a Dirichlet boundary condition imposes the ambient temperature~$T_{\mathrm{amb}}$ at the end of the spatial domain~$r_{\infty}$.
The boundary condition at the injection/extraction filter (i.e., $r=r_{0}$), however, depends on the operational mode of the ATES system, which justifies a piecewise system definition.
For each operational mode, a specific set of boundary conditions applies.
A Neumann boundary condition imposes a zero temperature gradient for extraction, e.g., for the warm ATES
\begin{equation*}
    \label{eq:boundary-condition-no-temp-gradient}
    \frac{\partial T_{\mathrm{w}}(r_{0}, t_{k})}{\partial r} = 0 \, .
\end{equation*}
The injection temperature is given by a Dirichlet condition.

For the MPC scheme, PDE~\eqref{eqn:T_PDE_nonlinear} is slightly modified, and the radial groundwater flow $v(r,t)$ is exchanged with the system input~$u(t)$, the volumetric pump flow between the warm and cold aquifer.
The relation is given by
\begin{equation*}
    v(r,t) = \frac{1}{2\pi r l \phi} u(t) \, ,
\end{equation*}
where $l$ reflects the length of the borehole's filter segment.
This equation holds for $r \in [r_{0}, r_{\infty}]$ and assumes that the borehole radius is larger than \qty{0}{\meter}.

The cocurrent HX is a central component ensuring the energy exchange between the ground and the building.
The outlet temperatures at the ATES side are given by \cite{Roetzel.2010},
\begin{equation}
    \label{eq:HX-state-dynamics}
    T_{\mathrm{c}}(r_{0},t_{k+1}) =  \left( 1 - \alpha_{\mathrm{a}}(t_{k}) \right) T_{\mathrm{w}}(r_{0},t_{k}) +  \alpha_{\mathrm{a}}(t_{k}) T_{\mathrm{r}}(t_{k}) \, ,
\end{equation}
where $\alpha_{\mathrm{a}}(t_{k})$ refers to the temperature change coefficient between the inlet and outlet at the ATES side of the heat exchanger.
This coefficient is a function of the system input $u(t_{k})$ and flow~$q_{\mathrm{b}}$ at the building side.
The outlet temperature at the building side~$T_{\mathrm{b}}(t_{k})$ is modeled accordingly.

Given the schematic illustration in Figure~\ref{fig:ATES-principle}, it is important to highlight that the HX's inlet at the ATES side is always connected to the storage where groundwater is extracted.
For heating, the warm ATES is connected, and for cooling, the cold ATES is connected.
Thus, the definition of the warm storage's borehole temperature~$T_{\mathrm{w}}(r_{0},t_{k})$ and cold storage's borehole temperature~$T_{\mathrm{c}}(r_{0},t_{k})$ depends on the operational mode, e.g.: while heating, $T_{\mathrm{w}}(\cdot)$ is defined by the storage temperatures and while cooling $T_{\mathrm{w}}(\cdot)$ is defined by the HX.

Following the modeling framework of \cite{vanRandenborgh.2024}, the ATES model $\ff_{\mathrm{ates}}(\xb(t_{k}), u(t_{k}), T_{\mathrm{r}}(t_{k}))$ is given by
\begin{equation} \label{eq:pwnl-ates-system-dyn}
    \ff_{\mathrm{ates}}(\cdot) =
    \begin{cases}
        \ff_{\mathrm{heating}}(\cdot) & \text{if } u(t_{k}) > 0\\
        \ff_{\mathrm{inactivity}}(\cdot) & \text{if } u(t_{k}) = 0\\
        \ff_{\mathrm{cooling}}(\cdot) & \text{if } u(t_{k}) < 0 \, ,
    \end{cases}
\end{equation}
where $\xb(t_{k})$ represents the concatenation of the temperature at the building side~$T_{\mathrm{b}}(t_{k})$ and the temperature profiles~$T_{\mathrm{w}}(\cdot)$ and $T_{\mathrm{c}}(\cdot)$ for the warm and cold ATES.
Note that in comparison to \eqref{eq:pwnl-ates-system-dyn} the partitions~$\Xbc_{i}$ of the domain only concentrate on the system input~$u$.
Furthermore, the functions ($\ff_{\mathrm{heating}}(\cdot)$, $\ff_{\mathrm{inactivity}}(\cdot)$, and  $\ff_{\mathrm{cooling}}(\cdot)$) are bi-linear in input~$u$ and state~$\xb$.
$T_{\mathrm{r}}(t_{k})$ is an exogenous input to the system as it represents the building's return temperature.

\section{THE NOVEL STATE OBSERVER}\label{sec:the-novel-state-observer}
The MHE model is a PWA approximation of the piecewise nonlinear surrogate model~\eqref{eq:pwnl-ates-system-dyn} and divides its domain into further partitions (of the system input~$u$).
Furthermore, it assumes a constant system input~$u_{i}$ for each partition~$\Xbc_{i}$ and avoids, consequently, bilinear state dynamics.
The accuracy of the MHE model can be arbitrarily increased by adding more partitions to the model.
For the MHE model, $s$ must be greater than three.

The adaptions yield the piecewise-affine state $\ff_{\mathrm{mhe}}(\xb(t_{k}), u(t_{k}), T_{\mathrm{r}}(t_{k}))$ and output dynamics $\hf_{\mathrm{mhe}}(\xb(t_{k}), u(t_{k}), T_{\mathrm{r}}(t_{k}))$,
\begin{subequations}\label{eq:pwa-mhe-model}
\begin{align}
    \ff_{\mathrm{mhe}}(\cdot) & = \begin{cases}
        \Ab_{1}(u_{1}) \xb(t_{k}) + \fb_{1}(\cdot) & \text{if } u(t_{k}) \in \mathbb{U}_{1}\\
        \vdots \\
        \Ab_{s}(u_{s}) \xb(t_{k}) + \fb_{s}(\cdot) & \text{if } u(t_{k}) \in \mathbb{U}_{s}
    \end{cases} \label{eq:mhe-state-dyn} \\
    \hf_{\mathrm{mhe}}(\cdot) & = \Cb \xb(t_{k}) + \Db u(t_{k}) + \eb(\cdot) \label{eq:mhe-output-dyn} \, ,
\end{align}
\end{subequations}
where $\mathbb{U}_{i} = \{u \, | \, \underline{u}_{i} \leq u \leq \overline{u}_{i}\} \, \forall i \in \{0, ..., s\}$ are polyhedral sets and $u_{i} \, \forall i \in \{0, ..., s\}$ is at the center of each partition~$\mathbb{U}_{i}$.
The state~$\fb_{i}(\cdot)$ and output~$\eb(\cdot)$ offset vectors depend on the exogenous input~$T_{\mathrm{r}}(t_{k})$, only.
Note that the system dynamics are autonomous in each partition~$\mathbb{U}_{i}$.
In the context of ATES systems, the output function~$\hf_{\mathrm{mhe}}(\cdot)$ is not piecewise because the sensor configuration, i.e., which states are measured, is independent of the partition~$\mathbb{U}_{i}$.
The system output~$\yb(t_{k})$ comprises the borehole temperatures, $T_{\mathrm{w}}(r_{0}, t_{k})$ and $T_{\mathrm{c}}(r_{0}, t_{k})$, and the outlet temperature at the building side of the heat exchanger~$T_{\mathrm{b}}(t_{k})$.
Additionally, the system input~$u(t_{k})$ and the building's return temperature~$T_{\mathrm{r}}(t_{k})$ are assumed known.
The output~$\Cb$ and feedthrough matrix~$\Db$ with the offset vector~$\eb(\cdot)$ are defined accordingly.

The equality constraints~\eqref{eq:mhe-op-constraints} of the MHE OP~\eqref{eq:MHE-OP} are updated with the MHE model~\eqref{eq:pwa-mhe-model}.
The fact that the partitions of the state dynamics $\ff_{\mathrm{mhe}}(\cdot)$ are only dependent on the system input $u$ allows to identify the correct partition~$\mathbb{U}_{i}$ based on the system input~$u$ and the exogenous input~$T_{\mathrm{r}}$.
With other words, a pre-processing step identifies the state matrix and state offset vector at each time~$t_{k} \, \forall k \in \{k^{\circ} - M, ..., k^{\circ}\}$ and makes the reformulation of \eqref{eq:MHE-OP} to an mixed-integer program obsolete.
In fact, the MHE OP~\eqref{eq:MHE-OP} becomes a quadratic program.

Outstandingly, the number of partitions~$s$ of the MHE model do not affect the computational cost of the quadratic program and can increase to infinity, as the state matrix $\Ab_{i}(u_{i})$ and state offset vector $\fb_{i}$ from \eqref{eq:pwnl-ates-system-dyn} can be directly computed by the pre-processing step.
In this case, the MHE model can perfectly imitate the surrogate ATES model~\eqref{eq:pwnl-ates-system-dyn}.

As a remark, it must be clarified that the MHE model may not be suitable for MPC, as increasing the number of partitions~$s$ increases the number of binary variables in the OCP and, thus, increases its computational effort.

Incremental observability of the MHE model~\eqref{eq:pwa-mhe-model} for a constant exogenous input~$T_{\mathrm{r}}$ can be verified with \cite[Alg. 1 \& 2]{Bemporad.2000}.
This requirement is necessary for a convergent MHE scheme \cite{FerrariTrecate.2002}.
Note that observability of each partition in \eqref{eq:mhe-state-dyn} is neither a necessary nor a sufficient condition for incremental observability (see, e.g., Chapter 4 in \cite{Bemporad.2000}).

\section{NUMERICAL STUDIES}\label{sec:numerical-studies}
The numerical studies are divided into three parts.
First, we will discuss how far states and their estimates have an impact on the MPC control actions, which, in fact, determine the upper bound of the spatial domain of the MHE model.
Next, the modeling accuracy of the MHE model is tested.
Finally, the state estimation performance of the novel MHE, the UKF from \cite{vanRandenborgh.2024}, and a linear time-varying Kalman filter (LTV-KF) is compared.

The study uses parameters of a real ATES system in Belgium.
A detailed description of the system is given by \cite{Vanhoudt.2011} and \cite{Desmedt.2007}.
Missing parameters (e.g., for the HX) are adequately chosen and disclosed in this paragraph.
The warm and cold ATES comprise rock and water with a porosity of~$\phi = 0.3$.
The volumetric heat capacity of the ATES is given by $c_{\mathrm{a}} = \qty{4.4625}{MJ.m^{-3}.K^{-1}}$, and the heat conduction coefficient~$\lambda$ is set to $\qty{3.5}{W.m^{-1}.K^{-1}}$ (for the MHE model only).
The filter length is~$l = \qty{38}{m}$, and its radius is $r_{0} = \qty{0.4}{m}$.
By assumption, the ambient temperature remains at $T_{\mathrm{amb}} = \qty{284.85}{K}$.
All models are temporally discretized with $\Delta t = \qty{3600}{s}$.
As the parameters of the building side are not disclosed by \cite{Desmedt.2007}, a constant flow rate~$q_{\mathrm{b}} = \qty{0.1}{m^{3}.s^{-1}}$ and building return temperatures for heating~$T_{\mathrm{r}} = \qty{276.15}{K}$ and cooling~$T_{\mathrm{r}} = \qty{291.15}{K}$ are assumed (given temperatures are after the APUs).
The upper and lower boundaries for the system input~$u(k)$ is set to $\pm \qty{0.0277}{m^3.s^{-1}}$.
All states of the cold and warm ATES are constrained by the ambient temperature~$T_{\mathrm{amb}}$ with the freezing point at $\qty{273.15}{K}$, and an upper bound~$\qty{293.15}{K}$ with the ambient temperature~$T_{\mathrm{amb}}$, respectively.
Further, to imitate a real-world system, subsurface parameters are perturbed spatially for the surrogate ATES model.
It is assumed that the heat conduction coefficient~$\lambda$ follows a uniform spatial distribution between upper \qty{5}{W.m^{-1}.K^{-1}} and lower \qty{3}{W.m^{-1}.K^{-1}} bounds.

\subsection{Estimation, but how far?}\label{sec:estimation-but-how-far}
In the context of ATES, MPC is typically used for heat demand reference tracking to achieve a comfortable indoor environment~\cite{vanRandenborgh.2024}.
The power production of the ATES~$P(\cdot)$ can be mathematically described by
\begin{equation*}
    P(t_{k}) = c_{\mathrm{w}} u(t_{k}) \left(T_{\mathrm{b}}(t_{k}) - T_{\mathrm{r}}(t_{k}) \right) \, ,
\end{equation*}
where $T_{\mathrm{b}}(\cdot)$ is directly depended on the HX state dynamics~\eqref{eq:HX-state-dynamics} and the extraction temperatures of the ATES.
Consequently, the OCP focuses on the groundwater's temperature close to the boreholes.
This leads to the question of whether states far from the borehole have any influence on the OCP's solution and therefore need not be estimated.

Given an MPC horizon of $N=12$ and a discretization step size of $\Delta t = \qty{3600}{s}$ the temperature of a particle at time~$t_{k^{\circ}}$ can maximally travel ($\|u\| = \qty{0.0277}{m^3.s^{-1}}$) by advection from the borehole $r_{0}$ to
\begin{equation}
    \label{eq:max-penetration-depth}
    \sqrt{\frac{N \, \Delta t \, u}{l \, \pi} + r_{0}^{2}} \approx \qty{3.2}{m} \, .
\end{equation}
Conversely, the temperature of the fluid with a distance of $\qty{3.2}{m}$ can be (theoretically) extracted within the MPC's prediction horizon~$N$ and thus have a potential influence on the power production~$P(\cdot)$ and the OCP's solution.

\begin{figure}
    \centering
    \includegraphics[trim={0.2cm, 5.9cm, 0.4cm, 0.35cm}, clip, width=\linewidth]{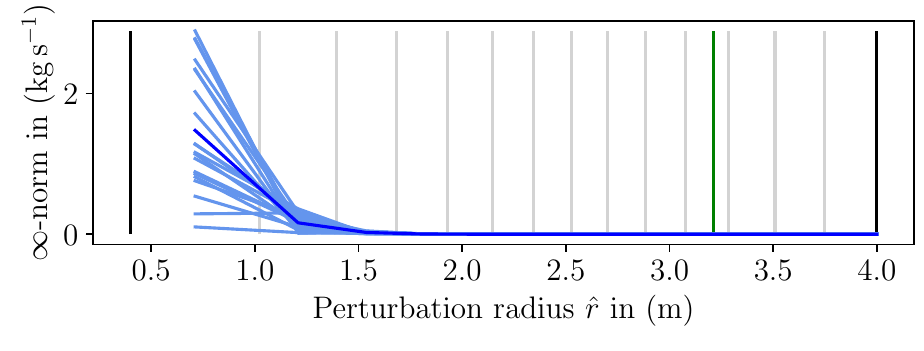}
    \caption{
        Infinity norm of the difference of the OCP's solution $\ub_{N}$ over the perturbation radius~$\rh$.
        Light blue lines indicate the results of single tests (20), whereas the dark blue line represents the mean.
        The black and gray vertical lines denote the bounds of the spatial domain and the mesh's cell boundaries.
        The green vertical line indicates the maximal temperature movement according to \eqref{eq:max-penetration-depth}.
    }
    \label{fig:mpc-state-sensitivity}
\end{figure}

The cost function for heat demand tracking is specified by
\begin{equation}
    \label{eq:mpc-sensitvity-cost-function}
    \Jf(\cdot) = \sum_{\kappa = k^{\circ}}^{k^{\circ}+N} \| D(t_{\kappa}) - P(t_{\kappa}) \|_{\Qb}^{2} + \| u(t_{\kappa}) \|_{\Rb}^{2} \, ,
\end{equation}
and the state dynamics~$\ff_{\mathrm{mpc}}(\cdot)$ are determined using the framework from \cite{vanRandenborgh.2024}.
Please note that the predicted energy demand of the building is given by $D(\cdot)$, and $\Rb$ and $\Qb$ represent appropriate weighting matrices.
The OPC~\eqref{eq:mpc} with cost function~\eqref{eq:mpc-sensitvity-cost-function} is a mixed-integer quadratic program.
For the numerical study, the solution of the OCP is compared using an initial state~$\xb(t_{k^{\circ}})$ and a perturbed state~$\xbh(t_{k^{\circ}}, \rh)$.
The perturbed state~$\xbh(\cdot, \rh)$ consists of states equal to the initial state (from $r_{0}$ to $\rh$) and perturbed states (between the perturbation radius~$\rh$ and the end of the spatial domain~$r_{\infty}$).
Perturbed temperatures are taken from a uniform distribution obeying the state constraints~$\mathbb{X}$.
The numerical study focuses on the difference of the OCP's solution, i.e., $\ub_{N}$, using the initial~$\xb(t_{k^{\circ}})$ and perturbed states~$\xbh(\cdot)$ over the perturbation radius~$\rh \in \left[ r_{0}, \qty{5}{m} \right]$.
Figure~\ref{fig:mpc-state-sensitivity} illustrates the results of the numerical study for $20$ repetitions.
The infinity norm of the difference of the OCP's solution~$\ub_{N}$ is displayed in light blue, whereas the average of all tests is given in dark blue.
In total, it can be seen that the OCP's solution alters only when the perturbation radius~$\rh$ is smaller than $\approx \qty{2}{m}$.
This indicates that \eqref{eq:max-penetration-depth} defines an appropriate upper bound.
As a consequence, states beyond this point do not influence the MPC control actions and may be neglected.
Equation~\eqref{eq:max-penetration-depth} assumes a ``worst-case'' scenario, where the system input is at its upper bound.
This is not experienced during the numerical study; the average system input~$u$ is $ \approx \qty{15}{kg.s^{-1}}$.
Based on this study's results, the following subsections have the end of the spatial domain at $r_{\infty} = \qty{4}{m}$.
The spatial domain of one ATES is discretized by $15$ cells, yielding $n = 33$ states.

\subsection{Accuracy of the moving horizon estimation model}\label{sec:accuracy-of-the-mhe-model}
\begin{figure}
    \centering
    \includegraphics[trim={0.8cm, 6.3cm, 0.4cm, 0.55cm}, clip, width=\linewidth]{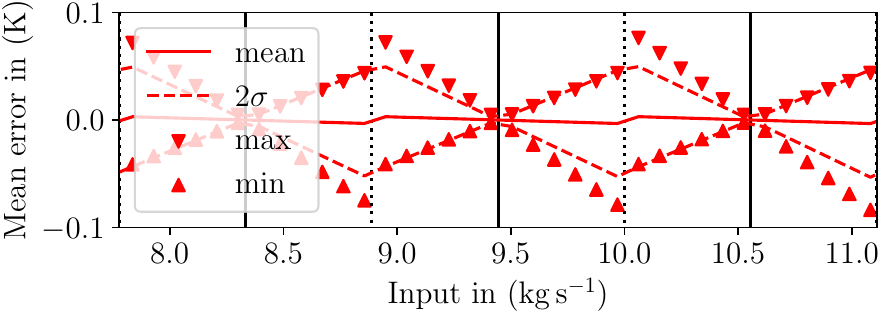}
    \caption{
        Illustration of the MHE model error (mean, 95\% confidence interval $2\sigma$, max, and min) in comparison to the surrogate ATES model.
        The vertical solid line corresponds to the center of each partition ($\mathbb{U}_{i}$), whereas the vertical dotted lines represent the bounds of each partition.
    }
    \label{fig:mhe-model-accuracy-local}
\end{figure}

Figure~\ref{fig:mhe-model-accuracy-local} illustrates the modeling error (mean, 95\% confidence interval, max, and min) of the MHE model~\eqref{eq:pwa-mhe-model} with $s=51$ partitions in comparison to the surrogate ATES model~\eqref{eq:pwnl-ates-system-dyn}.
The partitions of equal size cover the input constraints~$\mathbb{U}$ entirely.
The difference of the preceding state~$\xb(t_{k+1})$ using the MHE and surrogate ATES model for an identical initial state~$\xb(t_{k})$ over the system input~$u(t_{k})$ is determined.
As illustrated, the mean modeling error over all $n$ states behaves linearly within each partition~$\mathbb{U}_{i}$.
The maximal absolute error is given by \qty{0.147}{K}, and the standard deviation of the error is \qty{0.014}{K}.
For the following studies, the number of partitions $s$ is fixed, as the modeling error is considered sufficiently low.

\subsection{Comparison of novel MHE, UKF, and LTV-KF} \label{sec:comparison-of-novel-mhe-and-ukf}
The performance of the MHE, UKF, and LTV-KF is compared on synthetic data generated by simulating the surrogate model with an arbitrary initial state and input sequence.
By assumption, measurement noise follows a normal distribution with zero mean and a standard deviation of \qty{0.0333}{K}.
Process noise is given by a uniform distribution within a range of $\pm \qty{0.1}{K}$, i.e., $\mathbb{V} = \{ \nub \in \R^{n} | \, \| \nub \|_{\infty} \leq 0.1 \cdot \oneb_{n} \}$, where $\oneb_{n}$ is the 1-vector with $n$ entries.
In alignment with the PWA system model~\eqref{eq:pwa-mhe-model}, the states of the ATES system are measured at the outlet of the building side of the HX, and the last two sensors are located at the storages' boreholes, measuring injection and extraction temperatures.
Individual settings of the MHE, UKF, and LTV-KF are disclosed in the paragraphs below.
Each parameter is chosen carefully for a fair comparison of both estimators.

The time horizon of the MHE is $M = 40$.
For the MHE cost function, $Q = 10 \Ib_{n}$ and $R = 0.01 \Ib_{p}$, where $\Ib_{n} \in \R^{n}$ represents the identity matrix.
The initial penalty cost~$\Gamma_{t_{k^{\circ}-M}}(\cdot)$ is given by
\begin{equation*}
    \Gamma_{t_{k^{\circ}-M}}(\cdot) = \|\xb(t_{k^{\circ}-M}|t_{k^{\circ}-1}) - \xb(t_{k^{\circ}-M}|t_{k^{\circ}})\|_{\Sb}^{2}
\end{equation*}
and minimizes the $\Sb$-norm of the difference between the estimated state~$\xb(t_{k^{\circ}-M} | t_{k^{\circ}-1})$ at last time step and current time step~$\xb(t_{k^{\circ}-M} | t_{k^{\circ}})$.
For the numerical study, $\Sb = 0.001 \Ib_{n}$.
The process noise is constrained by $\mathbb{V}$.

Details of the UKF can be taken from \cite{vanRandenborgh.2024}.
Measurement and process noise are given by standard deviation~$\qty{0.0333}{K}$ with zero mean.
The standard deviation of the process noise reflects a distribution where \num{99.73}\% are contained by $\mathbb{V}$.

The LTV-KF exploits the fact that the non-linear surrogate ATES model~$\mathfrak{f}_{\mathrm{ates}}(\cdot)$ can be expressed by a linear time-varying system.
Details on the theory of LTV-KF are presented by, e.g., \cite{Chui.2017}.
The measurement and process noise are equivalently parameterized as in the paragraph above.

\begin{figure}
    \centering
    \includegraphics[trim={0.6cm, 6.3cm, 0.5cm, 0.5cm}, clip, width=\linewidth]{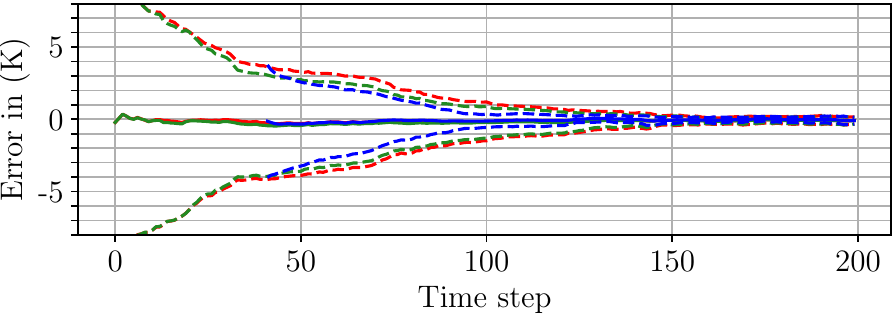}
    \caption{
        Illustration of the mean estimation error (solid line) for all $33$ states and its 95\% confidence interval (dashed line) for the MHE in blue, UKF in red, and LTV-KF in green.
        The MHE starts at $t_{40}$, which corresponds to the MHE's time horizon~$M$.
    }
    \label{fig:UKF-MHE-comparison}
\end{figure}

Figure~\ref{fig:UKF-MHE-comparison} shows the mean estimation error (solid line) for all estimated \num{33} states and its \num{95}\% confidence interval (dashed line) for the MHE in blue, UKF in red, and the LTV-KF in green.
At the beginning of the experiment ($t_{k} \leq t_{M}$), the MHE is not active because not enough output data is collected.
The UKF and LTV-KF start at $t_{1}$ and are initialized with a standard deviation of $\qty{0.4}{K}$ for the initial state, which is equal to $\zerob_{n}$ and implies that no prior knowledge about the system's state is given.
$\zerob_{n} \in \R^{n}$ refers to the 0-vector.
Whereas the first predictions of the UKF and LTV-KF have a low mean, the \num{95}\% confidence interval indicates a low reliability of the estimations.
As time proceeds, the 95\% interval tightens.
The MHE starts after $M$ time steps and the initial penalty cost function is initialized with the initial guess~$\xb(t_{0}|t_{M-1}) = \zerob_{n}$ (equal to the other estimators).
For comparison, the 95\% confidence interval of the estimators is (approximately) equal at this time.
After a few time steps, the MHE produces more confident state estimates, as its confidence interval is tighter than the UKF's and LTV-KF's.
The convergence rate of the MHE is greater than that of the other estimators.
Given the long operating hours of ATES systems, one can see that the 95\% confidence interval gets close to zero for longer simulation times, which proves the efficiency of this approach.
In general, the MHE keeps its estimates close to the simulated states and obeys the state constraints~$\mathbb{X}$.
In comparison, there is at least one UKF-estimated state every time step violating the state constraints~$\mathbb{X}$.
This is also experienced with the LTV-KF.

At $t_{75}$, it can be seen that the UKF has trouble estimating the states of the system and needs to regain its confidence.
These outliers may come from frequent changes between the operational modes of the ATES system, which cannot be detected immediately by the UKF and may destabilize the filter.
This may be linked to the ``short horizon syndrome'' of Kalman filters.

During the simulation, the mean estimation error of all estimators remains in the range of about $\qty{\pm 1}{K}$.

\section{CONCLUSION}\label{sec:conclusion}
ATES systems may play an important role in mitigating greenhouse gas emissions related to building climatization.
As literature indicates, a key challenge of ATES' worldwide deployment is its integration into the building's HVAC control, i.e., MPC.
By the nature of the application, state-based MPC schemes must deal with the scarcity of system outputs (measurements) as the placement of ground sensors necessitates large drilling costs.
To overcome this burden, state estimation may be used.
This, however, is challenging in regard to the piecewise nature of the system.
This work proposes a novel MHE scheme that avoids computationally expensive mixed-integer programming and exploits the piecewise nonlinear state dynamics of ATES systems.
With this, the MHE is based on an easy-to-solve quadratic program.
Concluding numerical studies examine the extent to which (spatial) distance ground temperature estimates and estimation errors influence the MPC's control actions.
Finally, the novel estimator is tested against a UKF.

\section{ACKNOWLEDGMENTS}

We thank Steffen Daniel for his review of the FVM.

\end{document}